\documentclass[%
 aip,
 jmp,%
 amsmath,amssymb,
 reprint,%
]{revtex4-1}

\usepackage{graphicx}% Include figure files
\usepackage{dcolumn}% Align table columns on decimal point
\usepackage{bm}% bold math
\begin{document}

\preprint{AIP/123-QED}

\title{Pressure dependence of the Verwey transition in magnetite: \\
an infrared spectroscopic point of view}

\author{J. Ebad-Allah}
\affiliation{Experimentalphysik 2, Universit\"at Augsburg, D-86195
Augsburg, Germany}

\author{L. Baldassarre}
\affiliation{Experimentalphysik 2, Universit\"at Augsburg, D-86195
Augsburg, Germany}

\author{M. Sing}
\affiliation{Physikalisches Institut, Universit\"at W\"urzburg, D-97074 W\"urzburg, Germany}

\author{R. Claessen}
\affiliation{Physikalisches Institut, Universit\"at W\"urzburg, D-97074 W\"urzburg, Germany}

\author{V. A. M. Brabers}
\affiliation{Physikalisches Institut, Universit\"at W\"urzburg, D-97074 W\"urzburg, Germany}

\author{C. A. Kuntscher}\email{christine.kuntscher@physik.uni-augsburg.de}
\altaffiliation{Department of Physics, Eindhoven University of Technology, 5600 MB Eindhoven, The Netherlands}

\date{\today}% It is always \today, today,

\begin{abstract}
We investigated the electronic and vibrational properties of magnetite at temperatures
from 300~K down to 10~K and for pressures up to 10~GPa by far-infrared reflectivity
measurements. The Verwey
transition is manifested by a drastic decrease of the overall reflectance
and the splitting of the phonon modes as well as the activation of additional
phonon modes. In the whole studied pressure range the down-shift of the overall reflectance spectrum saturates
and the maximum number of phonon modes is reached at a critical temperature,
which sets a lower bound for the Verwey transition temperature T$_{\mathrm{v}}$.
Based on these optical results a pressure-temperature phase diagram for magnetite is proposed.
\end{abstract}

%Use showkeys class option if keyword
\maketitle

\section{\label{sec:level1}Introduction}

Numerous experimental studies have been carried out on the
natural mineral magnetite (Fe$_{3}$O$_{4}$) to understand its puzzling
electronic and magnetic properties. It is the oldest known magnetic material and
the prototype material for the Verwey transition, which leads to charge ordering.\cite{Verwey39}
Fe$_{3}$O$_{4}$ has an inverse cubic spinel
structure at ambient conditions and is in a mixed-valence state
described as [Fe$^{3+}$]$_{\mathrm{A}}$[Fe$^{2+}$+Fe$^{3+}$]$_{\mathrm{B}}$O$_{4}$,
where A and B denote the tetrahedral and octahedral sites,
respectively, in the spinel structure AB$_{2}$O$_{4}$, with space group
Fd$\overline{3}$m.\cite{Sasaki97}
At ambient conditions magnetite can be classified as a bad metal
due to the presence of a small Drude contribution.\cite{Degiorgi87,Park98,Gasparov00}
With decreasing temperature magnetite undergoes a metal-to-insulator transition at the
so-called Verwey transition temperature T$_{\mathrm{v}}$$\approx$120 K,
concurrent with a lowering of the crystal structure symmetry from cubic
to monoclinic.\cite{Iizumi82,Wright01,Wright02,Senn12} Various experimental
investigations led to contradictory scenarios for the rearrangement of charges on the Fe sites
and the resulting charge ordering below T$_{\mathrm{v}}$.\cite{Nazarenko06,Kobayashi06,Pasternak03,Rozenberg07,Ovsyannikov08,Baudelet10}
In contrast, consistent results have been obtained by various groups\cite{Degiorgi87,Park98,Gasparov00}
regarding the changes in the optical properties when cooling below T$_{\mathrm{v}}$:
According to infrared reflectivity measurements as a function of temperature a clear opening of a charge gap ($\approx$0.14 eV) and the appearance of numerous infrared modes due to the crystal symmetry lowering were observed.

Besides the charge ordering pattern below T$_{\mathrm{v}}$, a
matter of controversy is furthermore the behavior of T$_{\mathrm{v}}$ as a function of
pressure. Resistivity measurements\cite{Todo01,Mori02} showed a sharp drop of T$_{\mathrm{v}}$
at the critical pressure $P_c$=7-8~GPa to zero and a metallization above $P_c$. This finding was confirmed lateron by ac magnetic susceptibility together with resistivity measurements.\cite{Spalek08} In Ref.\ \onlinecite{Spalek08}
furthermore the existence of a quantum critical point in the pressure-temperature phase diagram of magnetite was claimed.\cite{Spalek08}
In contradiction to these results, several groups found a linear decrease of T$_{\mathrm{v}}$ with increasing pressure based on resistivity measurements, with a linear pressure coefficient of either -2.8~K/GPa \cite{Kakudate79,Tamura90,Rozenberg96} or -5~K/GPa \cite{Samara68,Ramasesha94}. A recent Raman study confirmed the linear decrease of T$_{\mathrm{v}}$ with -5 K/GPa.\cite{Gasparov05}
The absence or presence of a quantum critical point could be related to the level of hydrostaticity in the pressure cell, which may vary for the different types of experiments. Also, the measurement technique might play a role.

The above-mentioned controversy regarding T$_{\mathrm{v}}$($P$) motivated us to study the Verwey transition in magnetite as a function of pressure by far-infrared reflectivity measurements, which comprise both electronic and vibrational properties.
The goal of our investigation is not primarily the precise determination of the pressure dependence of the Verwey transition temperature like in other studies, but to characterize the phases of magnetite close to the Verwey transition
from an infrared spectros\-copic point of view: The overall level of the reflectance spectrum and the phonon mode activation and splitting are considered for various temperatures and pressures.
Based on these two criteria we propose a pressure-temperature phase diagram. We also relate our results to those of earlier works.

\begin{figure}[t]
\includegraphics[angle=0,width=0.85\columnwidth]{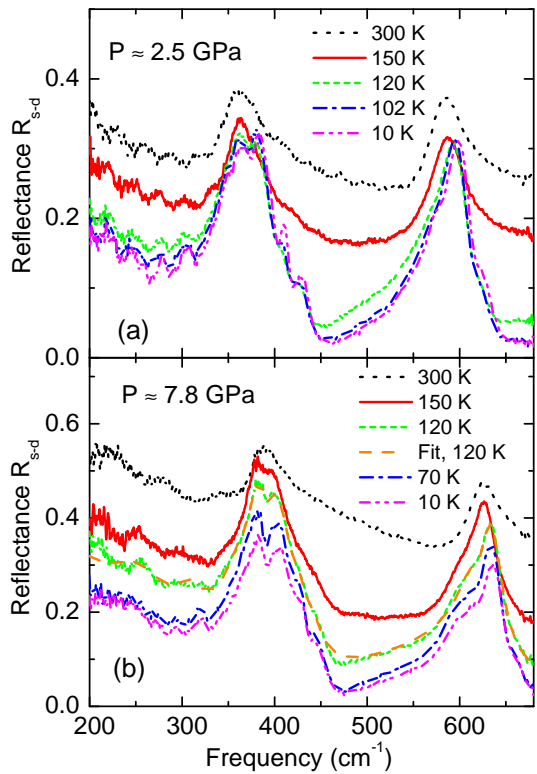}
\caption{Reflectance spectrum of magnetite for various temperatures at
around (a) 2.5 GPa and (b) 7.8 GPa. In (b) additionally the fitting curve
based on the Drude-Lorentz model for the reflectance spectrum at 120~K is shown.}
\label{fig:reflectance}
\end{figure}

\section{Methods}
The single crystals of magnetite used in this work were grown from polycristalline
Fe$_{3}$O$_{4}$ bars by using a floating-zone technique with
radiation heating.\cite{Brabers71} The polycrystalline bars were prepared from $\alpha$-Fe$_2$O$_3$
by the usual ceramic procedures as pressing and sintering in an adapted oxygen atmosphere.
The quality of the crystals was checked by electrical
transport measurements, showing a sharp increase of the resistivity
by a factor of about 100 at the critical temperature T$_\mathrm{v}\approx$
122 K, which is characteristic for the Verwey transition.\cite{Ebad-Allah12} 
The low-temperature reflectance
measurements under pressure were conducted in the frequency range
from 200 to 700 cm$^{-1}$ with a frequency resolution 1~cm$^{-1}$ partly at the infrared
beamline of the synchrotron radiation source ANKA and partly in
the lab at Augsburg university.
A clamp diamond anvil cell (Diacell cryoDAC-Mega)
equipped with type IIA diamonds, which
are suitable for infrared measurements, was used
for the generation of pressures up to 10~GPa. Finely ground CsI powder was
used as quasi-hydrostatic pressure transmitting medium, in order to ensure a well-defined sample-diamond interface throughout the experiment.
The pressure in the diamond anvil cell (DAC) was determined
{\it in situ} in the cryostat by the ruby luminescence method.\cite{Mao86}
The width of the phonon modes in the measured high-pressure reflectance spectra is comparable to that reported earlier on samples at ambient conditions,
\cite{Degiorgi87} evidencing the quasi-hydrostatic conditions in the DAC.

The reflectance measurements at low temperature
and high pressure were carried out using a home-built infrared
microscope coupled to the FTIR spectrometer and maintained
at the same vacuum conditions, in order to avoid
absorption lines of H$_2$O and CO$_2$ molecules. The infrared
radiation was focused on the sample by all-reflecting
Schwarzschild objectives with a large working distance of about
55~mm and 14x magnification. The DAC was mounted in a continuous-flow helium cryostat (Cryo Vac KONTI cryostat).
More details about the geometry of the reflectivity measurements can be found in our earlier
publications.\cite{Pashkin06,Kuntscher06} As reference, we used the intensity
reflected from the CuBe gasket inside the DAC. All reflectance spectra shown in this
paper refer to the absolute reflectance at the sample-diamond
interface, denoted as R$_{\mathrm{s-d}}$. Furthermore, corrections regarding
the decaying intensity of the synchrotron radiation with time have been taken
into account.

\begin{figure}[t]
\includegraphics[angle=0,width=0.9\columnwidth]{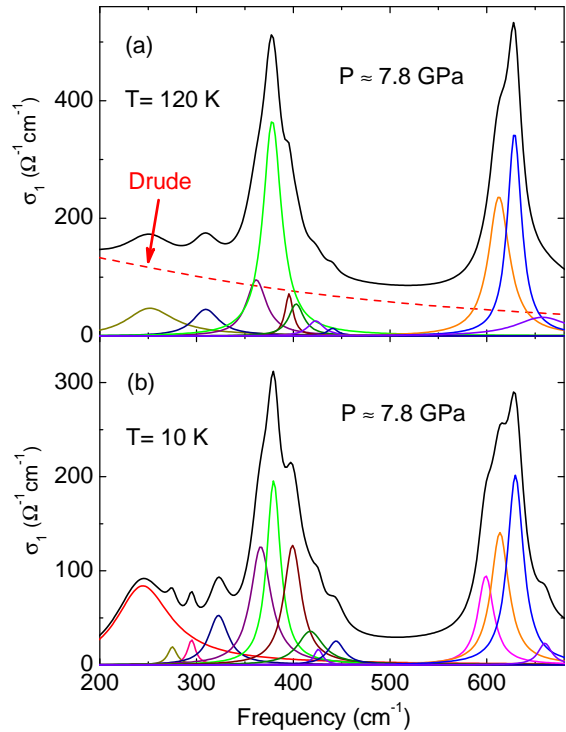}
\caption{Real part of the optical conductivity of magnetite at 7.8~GPa and (a) 120 K and (b) 10 K,
obtained from the fitting of the reflectance spectra with the Drude-Lorentz model.
The contributions (Lorentz oscillators; Drude term at
120~K) based on the fitting are also shown.}
\label{fig:conductance}
\end{figure}

\begin{figure}[t]
\includegraphics[angle=0,width=0.9\columnwidth]{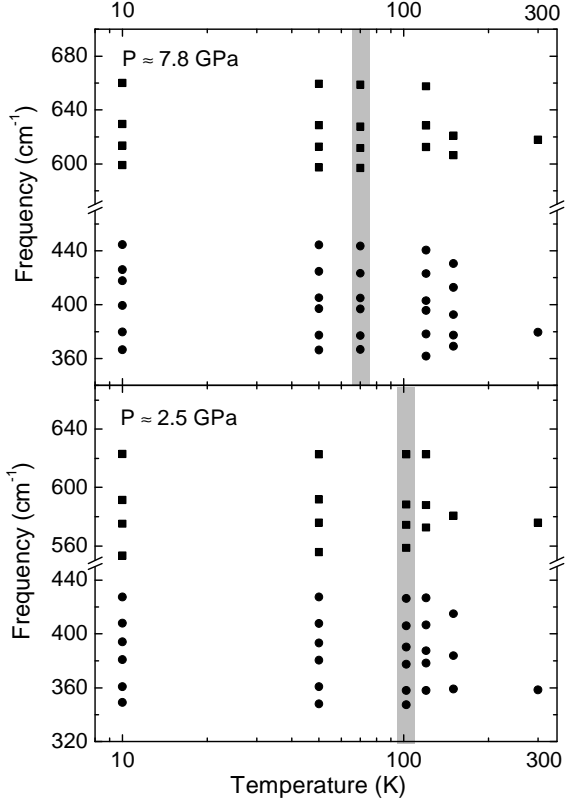}
\caption{Splitting of the two phonon modes and activation of phonon modes with decreasing
temperature at 2.5 and 7.8 GPa. At a critical temperature (marked by a gray bar) the maximum number of phonon modes is reached.}
\label{fig:phonon}
\end{figure}

\begin{figure}[h]
\includegraphics[angle=0,width=0.9\columnwidth]{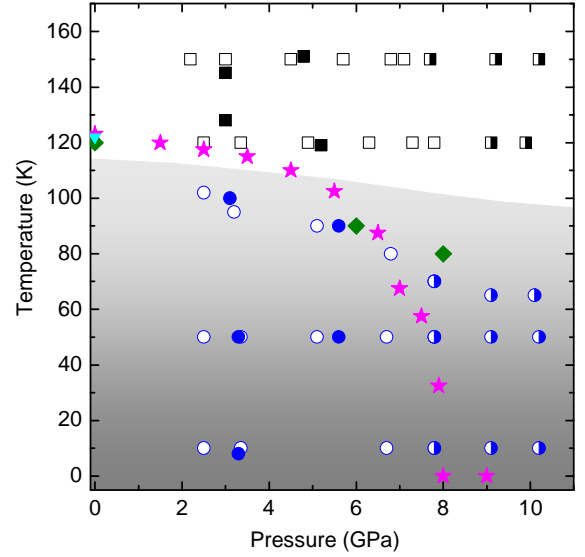}
\caption{Pressure-temperature phase diagram of magnetite for temperatures below 160~K
and pressures up to $\approx$10~GPa. The squares correspond to the ``bad metal'' state
and the circles to the insulating state. The results from three data sets are distinguished by
empty, filled, and half-filled symbols. The gray region indicates the charge-ordered, insulating phase as observed by our optical data. The Verwey transition temperature based on earlier dc transport measurements at ambient pressure is marked by the filled down triangle.\cite{Ebad-Allah12}
The results for the Verwey transition temperature T$_{\mathrm{v}}$ of Ref.\ \onlinecite{Mori02}
are marked by filled stars, and those of Ref.\ \onlinecite{Gasparov05} by filled diamonds.}
\label{fig:phase-diagram}
\end{figure}

\section{Results and Discussion}

The far-infrared reflectance spectrum of magnetite was measured for several pressures between
0 and 10~GPa
as a function of temperature. As an example, we show in Fig.\ \ref{fig:reflectance} (a) the reflectance spectrum
at $\approx$2.5~GPa for various temperatures.
At room temperature two oxygen phonon modes are observed with frequencies close to 355 cm$^{-1}$ and 565
cm$^{-1}$ at low pressure, consistent with earlier
reports.\cite{Degiorgi87,Park98,Gasparov00,Ebad-Allah09,Ebad-Allah12}
While cooling down from room temperature to 150~K, which is well above T$_{\mathrm{v}}$,\cite{Mori02,Spalek08,Kakudate79,Tamura90,Rozenberg96,Samara68,Ramasesha94,Gasparov05}
the overall reflectance spectrum gradually decreases [see Fig.\ \ref{fig:reflectance}(a)],
consistent with a ``bad metal'' behavior also found in temperature-dependent
dc resistivity data.\cite{Ebad-Allah12}
Below $\approx$150~K the overall reflectance
decreases drastically with a saturation below $\approx$102~K. Furthermore, below 150~K the
reflectance spectrum becomes more complex because of the phonon mode splitting and the activation
of additional low-frequency phonon modes.
By comparing the temperature dependence of the reflectance spectra for two pressures (2.5 and 7.8~GPa),
as depicted in Figs.\ \ref{fig:reflectance}(a) and (b), it is obvious that this
overall evolution of the reflectance spectrum with decreasing temperature is
not strongly dependent on the applied pressure. Furthermore, at a fixed temperature the overall reflectance spectrum is enhanced with increasing pressure, which is consistent with the pressure-induced increase of the conductivity according to dc transport measurements.\cite{Mori02}

For a quantitative characterization of the pressure- and temperature-induced
changes in the optical response, the reflectance spectra were fitted according to the Fresnel
equation for normal-incidence reflectivity taking into account the diamond-sample interface:
\begin{equation} \label{equ:fresnel}
R_{\mathrm{s-d}}=\left|\frac{n_{\mathrm{dia}}-\sqrt{\epsilon_{\mathrm{s}}}}{n_{\mathrm{dia}}+\sqrt{\epsilon_{\mathrm{s}}}}\right|^{2} ;~ ~ \epsilon_{\mathrm{s}}=\epsilon_{\infty}+\frac{i\sigma}{\epsilon_{0}\omega} \quad ,
\end{equation}
where $n_{dia}$ is the refractive index of diamond and assumed to be independent of pressure and temperature, and $\epsilon_{\mathrm{s}}$ is the complex dielectric function of the sample.
From the function $\epsilon_{\mathrm{s}}(\omega)$ the real part of the optical conductivity, $\sigma_{1}(\omega)$, can be calculated according to Eq.\ (\ref{equ:fresnel}), where
$\epsilon_{\infty}$ is the background dielectric constant (here $\epsilon_{\infty}\approx1$).
$\epsilon_{\mathrm{s}}(\omega)$ was assumed to follow the Drude-Lorentz model.\cite{Wooten72}
An example for the fitting is depicted in Fig.\ \ref{fig:reflectance}(a) for T=120~K at $P$=7.8~GPa: To obtain a good fit of the reflectance spectrum, we had to include
a Drude term and several Lorentz oscillators describing the phonon modes, while the
high-frequency extrapolation was modeled according to ambient-temperature data.\cite{Ebad-Allah12} It is important to note here that the extrapolations are not signi\-fi\-cant for our conclusions (see below) on the pressure-temperature phase diagram of magnetite.
The various excitations (Drude term, phonon modes) at T=120~K and 7.8~GPa are illustrated
in Fig. \ref{fig:conductance} (a) together with the real part of the optical conductivity within the measured frequency range.
At the lowest studied temperature (10~K) the optical conductivity can be described as a sum of
Lorentz oscillators reflecting the rich phonon spectrum below T$_{\mathrm{v}}$ [see Fig. \ref{fig:conductance} (b)].

As a criterium for entering the charge-ordered state one can take the saturation of the down-shift of the reflectance spectrum (see Fig.\ \ref{fig:reflectance}) and the absence of a Drude
term. Concomitant with the overall lowering of the reflectance spectrum occurs the
splitting of the phonon modes and the activation of additional modes. The mode splitting and activation are manifestations
of symmetry lowering of the crystal structure (from cubic to monoclinic) at the Verwey
transition.\cite{Degiorgi87,Park98,Gasparov00} Therefore, reaching the maximum number of phonon modes serves as an additional criterium for entering the charge-ordered phase.
The temperature-dependent phonon spectrum was characterized based on the Drude-Lorentz fitting, and
the extracted mode frequencies are shown in Fig.~\ref{fig:phonon}
at 2.5 and 7.8~GPa. Interestingly, the splitting and activation of phonon modes already
start at around 150~K, i.e., well above T$_{\mathrm{v}}$, in agreement with earlier infrared
studies.\cite{Gasparov00,Pimenov05}
The temperature at which a saturation regarding the down-shift of the reflectance spectrum and
the number of phonon modes is reached sets a lower bound for T$_{\mathrm{v}}$.

Based on two criteria - the down-shift of the reflectance spectrum and the phonon mode splitting/activation -
we can mark the ``bad metal'' and insulating phases in the pressure-temperature phase diagram
of magnetite, as shown in Fig.~\ref{fig:phase-diagram}. The most important result is that
we do not observe a suppression of the insulating state even at the highest pressure applied (10~GPa), in agreement with some earlier studies.
\cite{Kakudate79,Tamura90,Rozenberg96,Samara68,Ramasesha94,Gasparov05}
We do not observe the metallization of magnetite above $\approx$8~GPa and the occurrence of a quantum critical point, as claimed in other works.\cite{Todo01,Mori02,Rozenberg06,Spalek08}
Whether or not this is due to the quasi-hydrostatic conditions in the DAC remains an open question, as no systematic study on the pressure dependence of the Verwey transition temperature in the same sample for various pressure transmitting media exists.

\section{Conclusion}

In conclusion, based on the temperature- and pressure-dependent optical response of magnetite
we find the typical signatures of the Verwey transition, namely (i) the overall lowering of the
reflectance related to the entering of the charge-ordered state, and (ii) the splitting and activation of phonon modes due to the lowering of the crystal symmetry.
The downshift of the reflectance spectrum and the splitting/activation of phonon modes
are completed at a critical temperature, which sets a lower bound for T$_{\mathrm{v}}$.
Based on these results, we propose a phase diagram for magnetite showing the ``bad metal'' and
the insulating phases. The metallization of magnetite above $\approx$8~GPa
and the occurrence of a quantum critical point is not observed in our data.

\begin{acknowledgments}
We acknowledge the ANKA Angstr\"omquelle Karlsruhe for the provision
of beamtime and thank B. Gasharova, Y.-L. Mathis,
D. Moss, and M. S\"upfle for assistance using the beamline ANKA-IR.
We thank K. Syassen for providing valuable information about the
construction of the home-made infrared microscope.
This work was financially supported by the German Science Foundation (DFG)
through SFB 484.
\end{acknowledgments}

\end{document}